\documentclass[twoside]{mhd}
\usepackage{amsmath}
\usepackage{graphicx}
\usepackage{bm}
\mhdhead{40}{1}{1}	

\title{Transitions in a magnetized quasi-laminar\\ spherical Couette Flow}

\author{C.~Kasprzyk\inst{1*}, E.~Kaplan\inst{2}, M.~Seilmayer\inst{1}, F.~Stefani\inst{1}}

\institute{Helmholtz-Zentrum Dresden-Rossendorf, Bautzner Landstra{\ss}e 400, 01328 Dresden, Germany 
\and Universit\'e Grenoble Alpes, CNRS, 
ISTerre, CS 40700, F-38000 Grenoble Cedex 9, France}

\begin{document}
\maketitle


\begin{abstract}
First results of a new spherical Couette experiment are presented. The
liquid metal flow in a spherical shell is exposed to a homogeneous
axial magnetic field.  For a Reynolds number $\rm Re=1000$, we study
the effect of increasing Hartmann number $\rm Ha$.  The resulting flow
structures are inspected by ultrasound Doppler velocimetry.  With a
weak applied magnetic field, we observe an equatorially anti-symmetric
jet instability with azimuthal wave number $m=3$. As the magnetic
field strength increases, this instability vanishes. When the field is
increased further, an equatorially symmetric return flow instability
arises.  Our observations are shown to be in good agreement with
linear stability analysis and non-linear flow simulations.
\end{abstract}


\section*{Introduction}

The term spherical Couette flow refers to the fluid motion between two
rotating spherical shells.  If the liquid is electrically conducting
and exposed to an external magnetic field, the set-up is sometimes
called magnetized spherical Couette (MSC) flow.  For the special case
of resting outer cylinder, the system is completely defined by three
dimensionless parameters \cite{Ruediger_2013}: the Reynolds number
${\rm Re}=\Omega_i r_i^2/\nu$ as a measure of the rotation (with
angular velocity $\Omega_i$ of the inner sphere with radius $r_i$),
the Hartmann number ${\rm{Ha}}=B_0 r_i \sqrt{\sigma/\rho\nu}$ with
$B_0$ denoting the strength of the applied axial magnetic field, and
the geometric aspect ratio $\eta=r_i/r_o$ with the outer sphere radius
$r_o$. Here, $\nu$ denotes the kinematic viscosity, $\rho$ the
density, and $\sigma$ the conductivity of the fluid.

A long, albeit contentiously, discussed result of MSC flow was the
observation of an angular momentum transporting, magnetically induced
instability in a turbulent liquid sodium flow at $\rm Re \approx
10^7$, which was described in \cite{Sisan_2004} as the long
sought-after magnetorotational instability (MRI). In contrast to the
MRI as usually described \cite{Balbus_1991}, this instability was
non-axisymmetric and demonstrated an equatorial symmetry whose parity
depended on the strength of the applied magnetic field. Subsequent
numerical investigations \cite{Hollerbach_2009,Gissinger_2011} turned
up a collection of induction-free instabilities related to the
hydrodynamic jet instability, the Kelvin-Helmholtz-like Shercliff
layer instability, and a return flow instability that replicated the
parity properties, as well as the torque on the outer sphere (the
proxy measurement of angular momentum transport), as documented in
\cite{Sisan_2004}.

In order to study these types of instability in more detail, the new
apparatus HEDGEHOG ({\it H}ydromagnetic {\it E}xperiment with {\it
  D}ifferentially {\it G}yrating sph{\it E}res {\it HO}lding {\it
  G}aInSn) has been installed at Helmholtz-Zentrum Dresden-Rossendorf
(HZDR).  In contrast to \cite{Sisan_2004}, HEDGEHOG operates in a
quasi-laminar regime with two possible aspect ratios $\eta=0.35, 0.5$.
Numerical reference data is available for both these cases
\cite{Hollerbach_2009,Travnikov_2011}. Its typical operating
parameters, $\rm Re\sim 10^{3...4}$ and $\rm Ha\sim 10^{2...3}$, place
HEDGEHOG in a wider class of experiments dealing with inductionless
instabilities, such as the helical \cite{Stefani_2009} and azimuthal
MRIs \cite{Seilmayer_2014}, and the Tayler instability
\cite{Seilmayer_2012}.

\section{The experimental set-up}
\begin{figure}[h]
\includegraphics[width=0.99\textwidth]{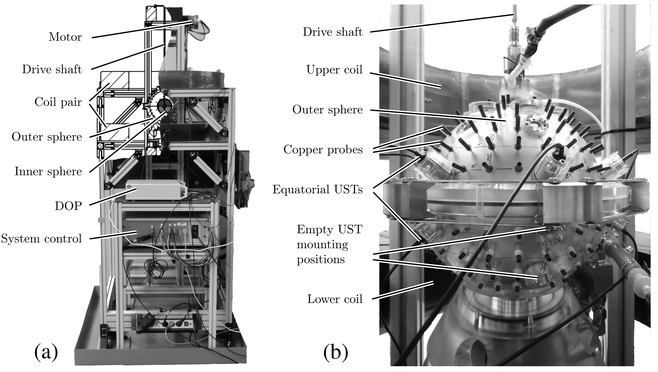}
\caption{The HEDGEHOG experiment: (a) Photograph of the facility. (b)
  Zoom on the spherical Couette module, showing the mountings for the
  Ultrasonic Doppler Transducers (UDT) and the copper electrodes for
  electric potential measurements.}
\label{fig:hedgehog_anlage}       
\end{figure}

Figure \ref{fig:hedgehog_anlage} gives an overview of the experimental
setup. The main parts, comprising the spherical vessel, the coil pair,
the system controls and the diagnostics are mounted on a rack shown in
the global view (Fig.~\ref{fig:hedgehog_anlage}a).

The central module (Fig.~\ref{fig:hedgehog_anlage}b) contains one of two
inner spheres ($r_i = 3$\,cm or 4.5\,cm) rotating in the center of an
outer sphere ($r_o=9$\,cm). Both spheres are made of Polymethyl
Methacetate (PMMA) acrylic with 30 cylindrical holders for ultrasonic
Doppler velocimetry (UDV), and 168 copper electrodes for electric
potential measurement. Two 90\,W motors drive the inner and outer
spheres independently of each other. The outer sphere rests on a
turntable connected to one motor; a 3\,mm drive shaft, 
which widens up to a 6\,mm outside the shell,
connects the inner sphere to the upper turntable. This 
paper is restricted to experiments with the outer sphere 
at rest and ${\rm Re} \le 10^4$ ($\Omega_i < 0.17$ Hz.).

The space between the two spheres is filled with GaInSn. Because of
the high density of this medium (6360\,kg/m\textsuperscript{3}), each
optional inner sphere contains a lead weight to counter the buoyancy
force. The axial magnetic field is provided by a pair of copper
electromagnets with central radii of 30\,cm, and a vertical gap 
of 31\,cm between them. This near Helmholtz configuration produces a static,
vertical magnetic field with less than one percent inhomogeneity in
the relevant measurement volume, 
providing ${\rm Ha} \le 160$ ($B \le 130$\,mT).

\begin{figure}[h]
\includegraphics[width=0.99\textwidth]{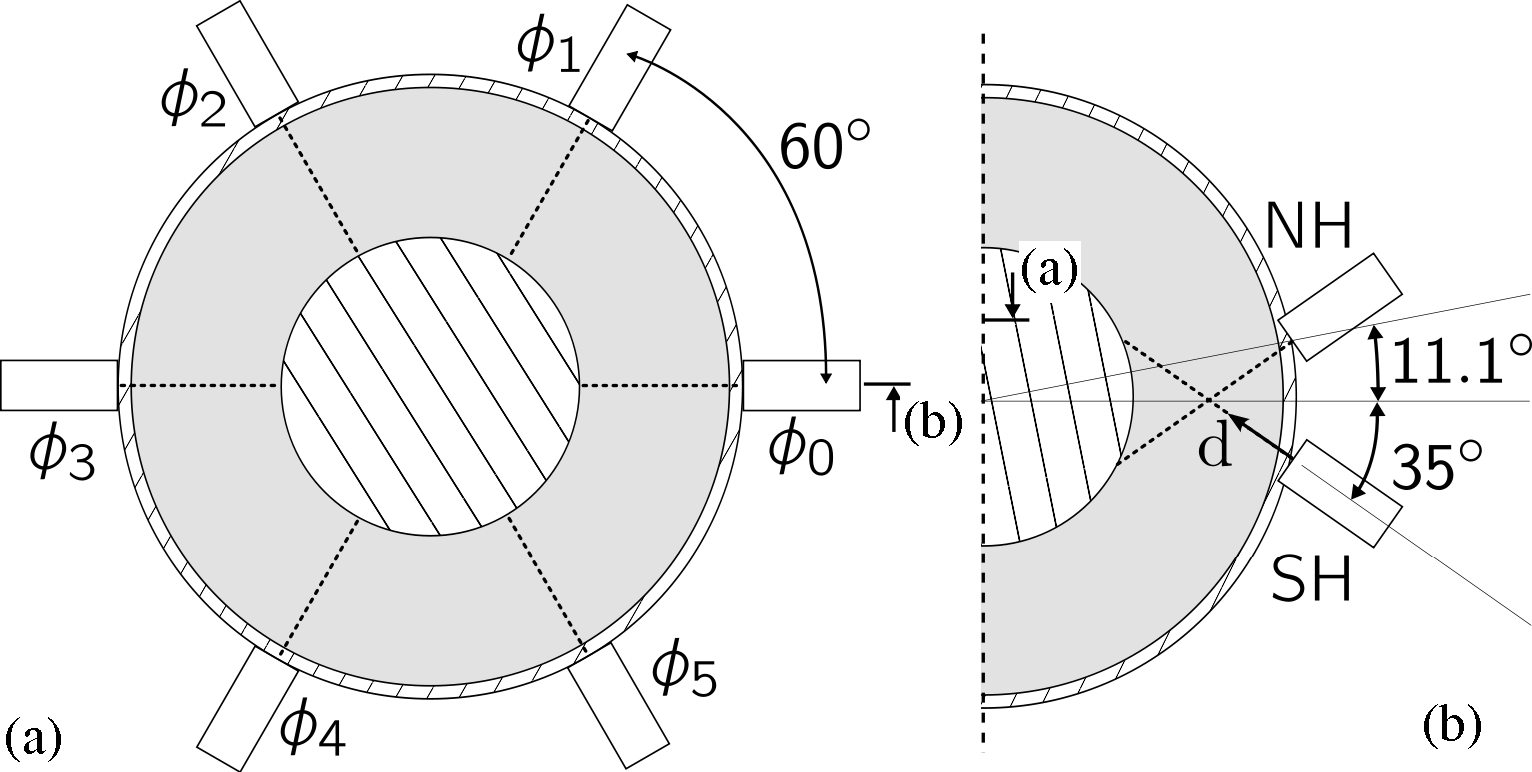}
\caption{UDV sensor configuration. (a) Polar view. (b) Meridional
  view. The NH and SH sensors are positioned at angles of $\pm
  11.1^{\circ}$ from the equator, Note, however, that their viewing
  angle has been chosen as $\pm 35^{\circ}$, in order to identify
  clearly the jet instability at low values of $\rm Ha$.}  %
\label{fig:sensor}
\end{figure}

Ultrasound transducers (USTs) mounted on the outer sphere allow a wide
variety of velocity field measurements from 30 different positions (5
different latitudinal positions repeated at 6 evenly spaced
longitudes, see Fig.~\ref{fig:sensor}).  The maximal wall thickness
between the UST and the fluid is $7$\,mm.  We use USTs with 4\,MHz
emitting frequency and the ultrasonic Doppler velocimeter (UDV)
DOP3010 (Signal Processing SA, Switzerland).  In
future a large number of installed copper pins will additionally be
used as electro-static potential probes.  The outer sphere and the
provided measurement technique are shown in Fig.
\ref{fig:hedgehog_anlage}b.

\section{Numerical predictions}

The experiment's dynamics are described by the incompressible Navier-Stokes equation,
\begin{align}
  \nabla \cdot \mathbf{U} &= 0 \\ \frac{\partial \mathbf{U}}{\partial
    t} &= \nabla p^* + {\rm Re} \left(\nabla \times \mathbf{U}\right)
  \times \mathbf{U} + {\rm Ha^2} \left(\nabla \times \mathbf{B}\right)
  \times \mathbf{B}
\end{align}
\noindent where $p^*$ is a reduced pressure absorbing all potential
forces, and no-slip conditions at the inner and outer spheres define
the drive of the flow by the rotating inner sphere. The Navier stokes
equation couples to the magnetic induction equation in its
inductionless limit,
\begin{equation}
  0 = \nabla^2 \mathbf b + \nabla \times \left(\mathbf{U} 
  \times \mathbf{B}_0\right) \label{eqn:Induction}.
\end{equation}
\noindent  In this limit the generation of a
magnetic field $\mathbf{b}$ from the interaction of a velocity field
$\mathbf{U}$ with a background field $\mathbf{B}_0$ is exactly
balanced by its diffusion. Initial simulations of the HEDGEHOG were
carried out with the code described in \cite{Hollerbach_2000}. These
provided predictions of the diagnostic outputs and demonstrated a
possible saturation mechanism for the instabilities
\cite{Kaplan_2014}. Further simulations were carried out with 
an adapted version of the 
MagIC code \cite{Magic_2016}, which has a long and very successful
record of simulating dynamos in spherical geometry \cite{Wicht_2014,Wicht_2010}.

\begin{figure}[h]
\includegraphics[width=0.99\textwidth]{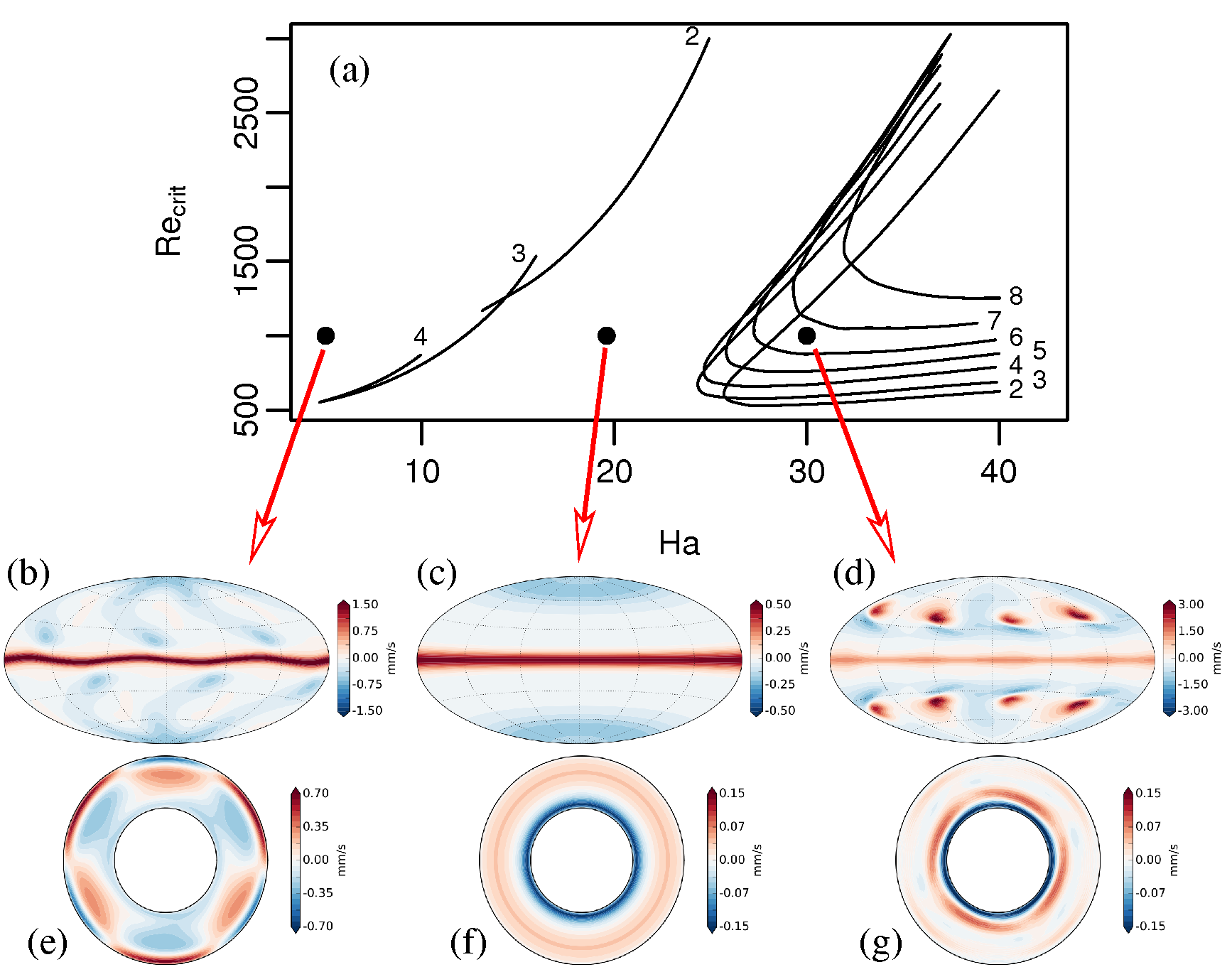}
\caption{Numerical simulations of the HEDGEHOG experiment.  (a)
  Instability boundaries in dependence on $\rm Ha$ and $\rm Re$,
  adopted from \cite{Travnikov_2011} and benchmarked with a linearized
  Navier-Stokes equation analysis of the axisymmetric base flows
  \cite{Hollerbach_2000,Hollerbach_2009}.  Left column: Instability at
  $\rm Rm=1000$ and $\rm Ha=5$.  (b) Meridional view of the simulated
  radial velocity component at $r=0.85 r_o$, showing the equatorially
  anti-symmetric jet instability. (e) Polar view of the meridional
  velocity component at $\theta=\pi/2$ (the equatorial plane),
  indicating an $m=3$ azimuthal dependence.  Middle column: Stable
  base flow at $\rm Rm=1000$ and $\rm Ha=19.6$.  (c) Meridional view
  of the simulated radial velocity component at $r=0.85 r_o$. (f)
  Polar view of the meridional velocity component at $\theta=\pi/2$.
  Right column: Instability at $\rm Rm=1000$ and $\rm Ha=30$.  (d)
  Meridional view of the simulated radial velocity component at $r=0.6
  r_o$, showing the equatorially symmetric character of the return
  flow instability.  (g) Polar view of the meridional velocity
  component at $\theta=\pi/2$, indicating an $m=4$ azimuthal
  dependence.  }
\label{fig:travnikov}       
\end{figure}

Figure~\ref{fig:travnikov} shows the stability boundaries and
illustrates the typical instabilities for a MSC flow with aspect ratio
$\eta=0.5$.  First, Fig.~\ref{fig:travnikov}a shows the boundaries for
the different types of instability in the ${\rm Ha}-{\rm Re}$ plane.
The lines were adapted from \cite{Travnikov_2011}, although large
parts of the diagram have been reproduced by our own simulations. The
left and right full circles represent two distinct types of
instability which are separated by a region of stability, represented
by the middle full circle.  The spatial character of the flow at these
three points, as simulated by the MagIC code, is illustrated in
Fig.~\ref{fig:travnikov}b-g.

At low $\rm Ha=5$, the instability arises in the jet connecting the
equators of the inner and outer spheres, see
Fig.~\ref{fig:travnikov}b,d.  It is anti-symmetric with respect to the
equator, and characterized by an $m=3$ azimuthal dependence.  At $\rm
Ha=19.6$, the flow has completely re-stabilized.  At higher ${\rm
  Ha=30}$, another type of instability arises in the return flow, the
meridional circulation through the interior of the flow that connects
the equatorial jet to the pole , see Fig.~\ref{fig:travnikov}d,g.  It
is symmetric with respect to the equator, and has an $m=4$ azimuthal
dependence.

\section{Experimental results}

Before any further evaluation, intensive post processing of the raw
UDV data was necessary.  The chosen pulse repetition frequency and the
internal correlation algorithm cause the appearance of a broad
velocity range in the raw data.  Moreover, the acquired data contain
divers unphysical spikes, reaching values ten times higher than
physically expected.  Therefore, a reasonable velocity range based on
$r_i \Omega$ was pre-selected, with velocities out of this range being
skipped and replaced with linearly interpolated values from valid
neighbouring points. The cleaned velocity is resampled onto a uniform
grid and, eventually, smoothed by a Gaussian filter.

The data presented in the following is based on the velocity acquired
from six equally spaced UDV sensors on the northern hemisphere (NH),
and from one UDV sensor on the southern hemisphere (SH)
(Fig.~\ref{fig:sensor}).  Using only 6 sensor data in azimuthal
direction, a maximum wave number of $m=3$ can be resolved according to
the Nyquist-Shannon criterion (this sub-sampling leads to an ambiguity
in the identification of the $m=4$ instability, as discussed below).
The dominant frequency of the equatorially symmetric and
anti-symmetric velocity parts can be inferred from spectrograms. The
data from the facing NH and SH UDV pair are  decomposed into
equatorially symmetric and anti-symmetric components ($u_s =
(u_{\rm{NH}}+u_{\rm{SH}})/2$ and $u_a = (u_{\rm{NH}}-u_{\rm{SH}})/2$,
respectively).

In the following we will describe three experimental runs, 
all done with a rotation rate of the inner sphere of 
0.027\,Hz, which amounts to 
$\rm Re=1000$. The currents in the Helmholtz coils were chosen as
12.08\,A, 47.34\,A, 72.47\,A, corresponding to $\rm Ha=5,19.6,30$, 
respectively.

\begin{figure}[h]
\includegraphics[width=0.99\textwidth]{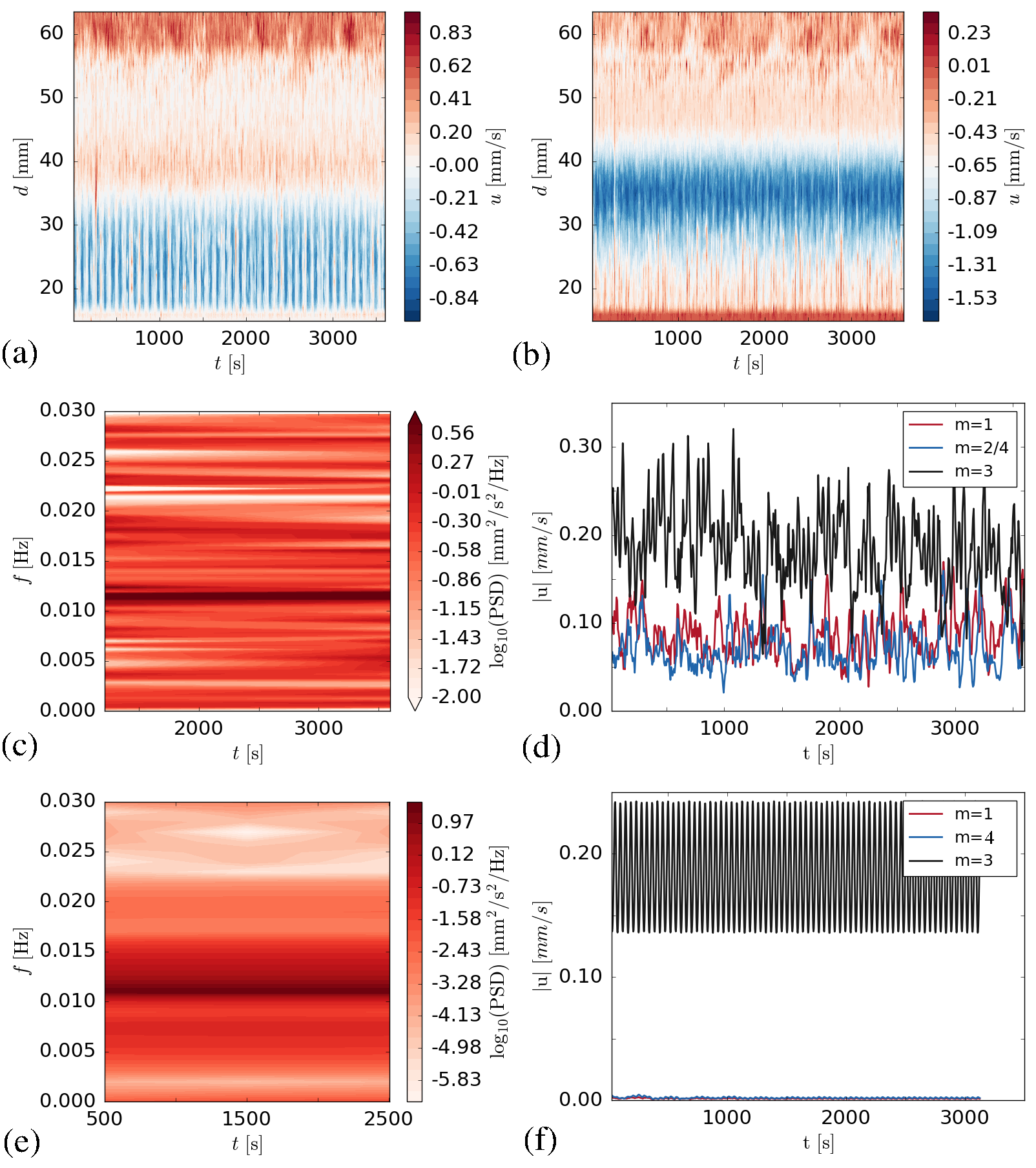}
\caption{Experimental and numerical data for $\rm Re=1000$ and $\rm
  Ha=5$.  (a) Anti-symmetric part $(u_{\rm{NH}}-u_{\rm{SH}})/2$ of the
  UDV measured velocity, showing a clear oscillation in the outer part
  of the spherical shell, close to the equator. (b) Symmetric part
  $(u_{\rm{NH}}+u_{\rm{SH}})/2$, indicating the concentrated jet at
  the equator.  (c) PSD of the anti-symmetric part, taken at a UDV
  beam depth of $d=25$\,mm, showing a clear peak around 0.011 Hz. (e)
  The same PSD, but simulated with MagIC.  (d) Azimuthal Fourier
  components of the UDV measured velocity taken at $d=25$\,mm, with a
  clearly dominating $m=3$ mode.  (f) The same for the simulated
  velocity, with a nearly pure $m=3$ mode.  }
\label{fig:daten_ha5}       
\end{figure}

Starting at the low value $\rm Ha=5$ (see the left point and column in
Fig.~\ref{fig:travnikov}), Fig.~\ref{fig:daten_ha5} reveals the
instability of the radial equatorial jet which was already known from
the purely hydrodynamic case \cite{Hollerbach_2006}.  Figure
\ref{fig:daten_ha5}a shows the equatorially anti-symmetric part of the
UDV measured velocity, taken at a UDV beam depth $d=25$\,mm. The
corresponding symmetric part is added in Fig.~\ref{fig:daten_ha5}b.
While the latter clearly shows a strong radial equatorial jet centered
around $d=35$ mm, the former shows the anti-symmetric instability
slightly above and below the equator. The dominant frequency of this
instability, $f\approx 0.011$ Hz, is derived from the PSD of the
experimental data (Fig.~\ref{fig:daten_ha5}c). This corresponds well
with the frequency computed by the MagIC simulations
(Fig.~\ref{fig:daten_ha5}e).  The azimuthal dependence of the
instability can be inferred from Fig.~\ref{fig:daten_ha5}d. At every
instant in time it provides the spectral content of the first
dominating azimuthal modes $m=1,2,3$. Note, that due to the
Nyquist-Shannon theorem we cannot distinguish between an $m=2$ and
an $m=4$ mode, which will become relevant below.  Here, the
$m=3$ mode is clearly dominant, which confirms the numerical
prediction as shown in Fig.~\ref{fig:daten_ha5}f.

\begin{figure}[h]
\includegraphics[width=0.99\textwidth]{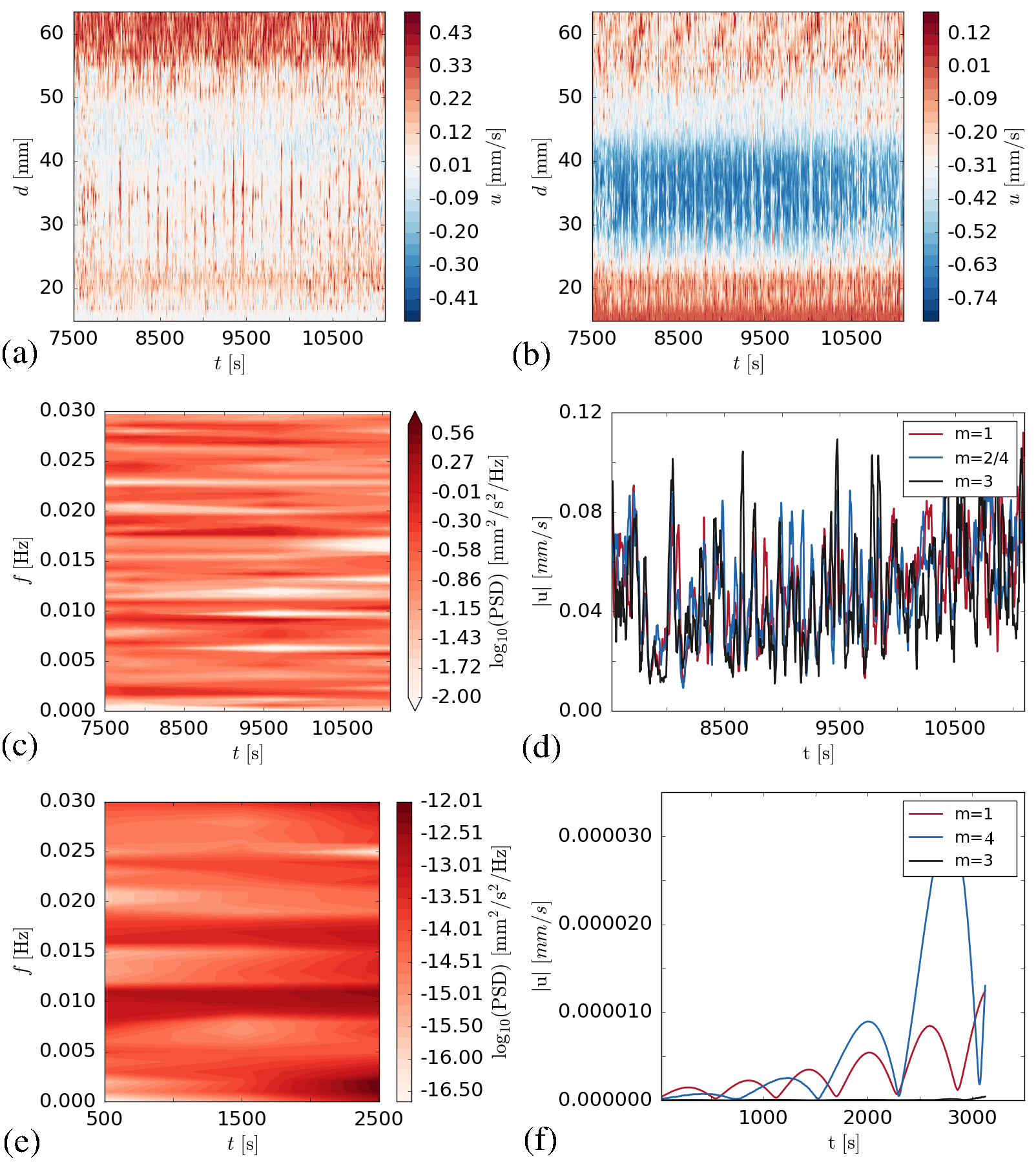}
\caption{Experimental and numerical data for $\rm Re=1000$ and $\rm
  Ha=19.6$.  (a) Anti-symmetric part $(u_{\rm{NH}}-u_{\rm{SH}})/2$ of
  the UDV measured velocity.  (b) Symmetric part
  $(u_{\rm{NH}}+u_{\rm{SH}})/2$.  (c) PSD of the anti-symmetric part,
  taken at $d=25$\,mm. (d) The same PSD, but simulated with MagIC.  (d)
  Azimuthal Fourier components of the UDV measured velocity taken at
  $d=25$\,mm.  (f) The same for the simulated velocity.  }
\label{fig:daten_ha19}       
\end{figure}

We switch now to the case $\rm Ha=19.6$ which lies in the isthmus of
stability shown in Fig.~\ref{fig:travnikov}) (middle column).  The
first observation to make is a significantly weakened (and slightly
broadened) radial jet (see Fig.~\ref{fig:daten_ha19}b, and notice the
changed colour scale of the velocity compared with that of
Fig.~\ref{fig:daten_ha5}b).  This weakened jet is no longer unstable
(Fig.~\ref{fig:daten_ha19}b). Thus, the PSD of the data, as
well as the Fourier components of the different $m$ modes, are much
reduced compared with those in Fig.~\ref{fig:daten_ha5}.

\begin{figure}[h]
\includegraphics[width=0.99\textwidth]{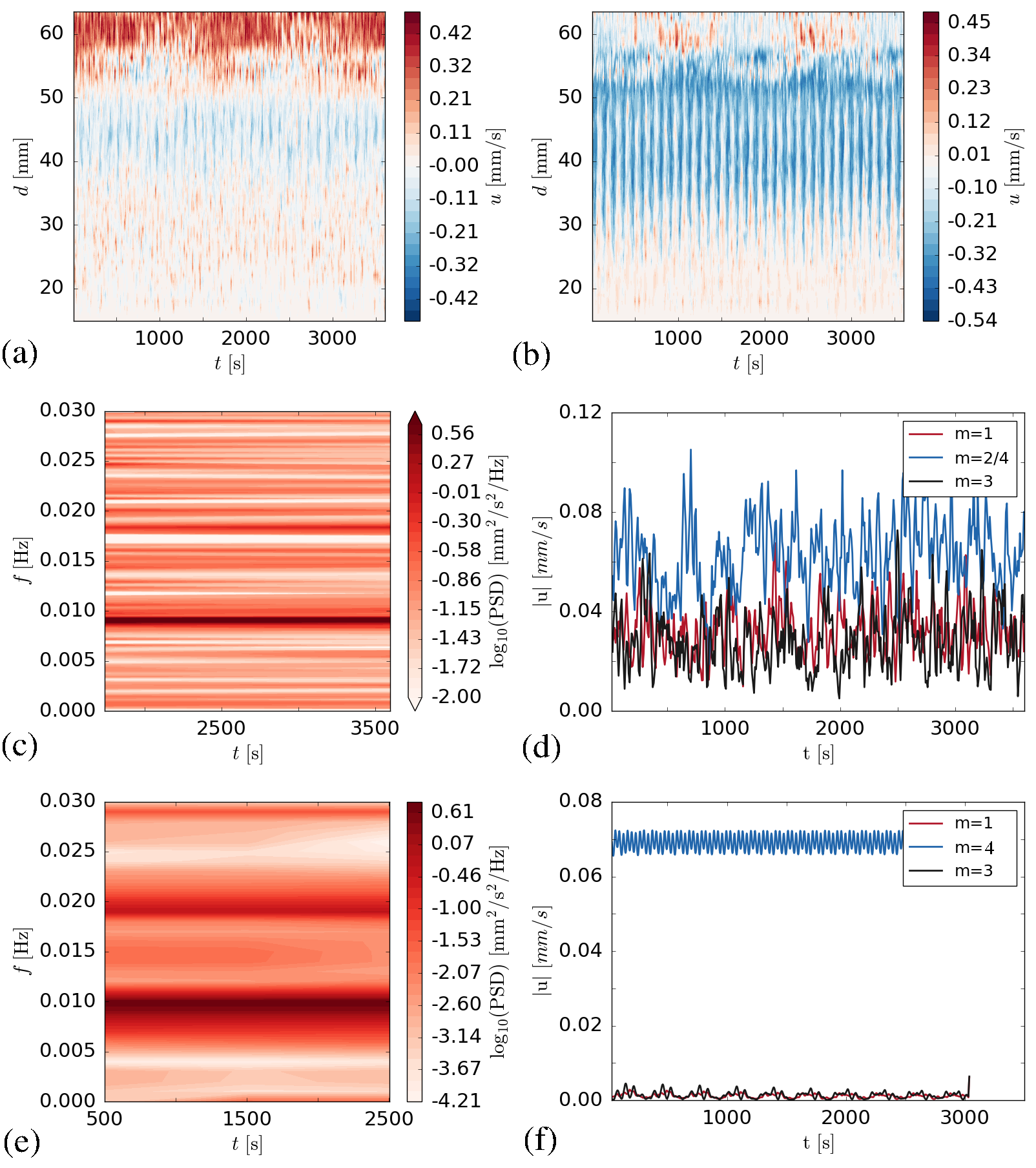}
\caption{Experimental and numerical data for $\rm Re=1000$ and $\rm
  Ha=30$.  (a) Anti-symmetric part $(u_{\rm{NH}}-u_{\rm{SH}})/2$ of
  the UDV measured velocity, showing a vanishing oscillation in the
  outer part of the spherical shell (corresponding to low beam depth
  $d$). (b) Symmetric part $(u_{\rm{NH}}+u_{\rm{SH}})/2$, indicating
  a clear oscillation at larger $d$, close to the tangential cylinder.
  (c) PSD of the symmetric part, taken at $d=37$\,mm, showing a clear
  peak around 0.009 Hz. (e) The same PSD, but simulated with MagIC.
  (d) Azimuthal Fourier components of the UDV measured velocity taken
  at a depth of 37\,mm, with dominating $m=2/4$ mode (not
  distinguishable due to Nyquist theorem).  (f) The same for the
  simulated velocity, with a nearly pure $m=4$ mode.  }
\label{fig:daten_ha30}       
\end{figure}

At $\rm Ha=30$ the instability is distinctly shifted towards the inner
sphere and acquires an equatorially symmetric character, which
indicates the onset of return flow instability.  Indeed, the
anti-symmetric signal (Fig.~\ref{fig:daten_ha30}a) is much weaker than
that in Fig.~\ref{fig:daten_ha5}a, while the symmetric part
(Fig.~\ref{fig:daten_ha30}b) is significantly stronger than in
Fig.~\ref{fig:daten_ha5}b.  Figure ~\ref{fig:daten_ha30}c shows the
PSD of the {\it symmetric} part, this time taken at a beam depth of
$d=38$ mm, with a strong peak at $f=0.009$ Hz, which is also
numerically found (Fig.~\ref{fig:daten_ha30}e).  Combined with this
agreement of the frequencies, the numerically determined dominant
$m=4$ mode (Fig.~\ref{fig:daten_ha30}f) justifies to resolve the
experimental ambiguity between the $m=2$ and the $m=4$ mode in favour
of the latter.

\section{Conclusions and outlook}

In this paper, we have described in detail the new MSC flow experiment
HEDGEHOG.  The azimuthal wave numbers and frequencies of the various
instabilities for increasing $\rm Ha$ turned out to be in good
agreement with linear and nonlinear numerical predictions.  In future,
electric potential measurements might help to further reduce the
remaining ambiguities with respect to the azimuthal wave number of the
observed modes.

\section*{Acknowledgments}
This research was supported by Deutsche Forschungsgemeinschaft (DFG)
under grant STE 991/1-1. We gratefully acknowledge many discussions
with Rainer Hollerbach, who also inspired us to set-up the
experiment. We thank Benjamin Gohl for his help in constructing the
apparatus, and Johannes Wicht for sharing with us the MagIC code.



\begin{thebibliography}{50}

\bibitem{Ruediger_2013}
{\sc G.~R\"udiger, L.~L. Kitchatinov, and R.~Hollerbach}.
\newblock {\it Magnetic Processes in Astrophysics\/} (Wiley-VCH, 2013).

\bibitem{Sisan_2004}
{\sc D.~R. Sisan, et~al.}
\newblock Experimental observation and characterization of the
  magnetorotational instability.
\newblock {\it Phys. Rev. Lett.\/}, vol.~93 (2004), Art. No.~114502.

\bibitem{Balbus_1991}
{\sc S.A. Balbus and J.F.   Hawley}.
\newblock {A powerful local shear instability in weakly magnetized disks. 
1. Linear analysis}.
\newblock {\it {Astrophys. J.}\/}, vol.~376 (1991), pp.~214--222.

\bibitem{Hollerbach_2009}
{\sc R.~Hollerbach}.
\newblock Non-axisymmetric instabilities in magnetic spherical Couette flow.
\newblock {\it Proceedings of the Royal Society of London A: Mathematical,
  Physical and Engineering Sciences\/}, vol.~465 (2009), pp.~2003--2013.

\bibitem{Gissinger_2011}
{\sc C.~Gissinger, H.~Ji, and J.~Goodman}.
\newblock Instabilities in magnetized spherical Couette flow.
\newblock {\it Phys. Rev. E\/}, vol.~84 (2011), Art. No.~026308.

\bibitem{Travnikov_2011}
{\sc V.~Travnikov, K.~Eckert, and S.~Odenbach}.
\newblock Influence of an axial magnetic field on the stability of spherical
  Couette flows with different gap widths.
\newblock {\it Acta Mechanica\/}, vol.~219 (2011), pp.~255--268.

\bibitem{Stefani_2009}
{\sc  F. Stefani et al}.
\newblock {Helical magnetorotational instability in 
a Taylor-Couette flow with strongly reduced Ekman pumping}.
\newblock {\it {Phys. Rev. E}\/}, vol.~97 (2006), Art. No. 184502.


\bibitem{Seilmayer_2014}
{\sc  M. Seilmayer et al}.
\newblock {Experimental evidence for non-axisymmetric 
magnetorotational instability in an azimuthal magnetic field }.
\newblock {\it {Phys. Rev. Lett.}\/}, vol.~113 (2014), Art. No. 024505.

\bibitem{Seilmayer_2012}
{\sc  M. Seilmayer et al}.
\newblock {Experimental evidence for a transient Tayler instability 
in a cylindrical liquid-metal column}.
\newblock {\it {Phys. Rev. Lett.}\/}, vol.~108 (2012), Art. No. 024501.

\bibitem{Hollerbach_2000}
{\sc R.~Hollerbach}.
\newblock {A spectral solution of the 
magneto-convection equations in spherical geometry.}.
\newblock {\it {Int J. Numer. Meth. Fluids}\/}, vol.~32 (2002), 
pp.~773--797.

\bibitem{Kaplan_2014}
{\sc E.~Kaplan}.
\newblock {Saturation of nonaxisymmetric 
instabilities of magnetized spherical Couette flow}.
\newblock {\it {Phys. Rev. E}\/} vol.~89 (2012), Art. No. 063016.

\bibitem{Magic_2016}
{\sc https://github.com/magic-sph/magic}.

\bibitem{Wicht_2014}
{\sc J.~Wicht}.
\newblock {Flow instabilities in the wide-gap spherical Couette system}.
\newblock {\it {J. Fluid Mech.}\/}, vol.~738 (2014), 
pp.~184--221.


\bibitem{Wicht_2010}
{\sc J.~Wicht and A.~Tilgner}.
\newblock {Theory and modeling of planetary dynamos}.
\newblock {\it {Space Sci. Rev.}\/}, vol.~152 (2010), 
pp.~501-542.



\bibitem{Hollerbach_2006}
{\sc R.~Hollerbach, M.~Junk, and C.~Egbers}.
\newblock Non-axisymmetric instabilities in basic state 
spherical Couette flow.
\newblock {\it Fluid Dynamics Research\/}, vol.~38 (2006), pp.~257--273.



\end{thebibliography}

\newcommand{\noopsort}[1]{}

\lastpageno

\end{document}